\documentclass[11pt,a4paper]{article}

\usepackage{fullpage}
\usepackage{setspace}
\usepackage{parskip}
\usepackage{titlesec}
\usepackage[section]{placeins}
\usepackage{xcolor}
\usepackage{breakcites}
\usepackage{lineno}
\usepackage{hyphenat}

\PassOptionsToPackage{hyphens}{url}
\usepackage[colorlinks = true,
            linkcolor = blue,
            urlcolor  = blue,
            citecolor = blue,
            anchorcolor = blue]{hyperref}
\usepackage{etoolbox}
\makeatletter
\patchcmd\@combinedblfloats{\box\@outputbox}{\unvbox\@outputbox}{}{%
  \errmessage{\noexpand\@combinedblfloats could not be patched}%
}%
\makeatother

\renewenvironment{abstract}
  {{\bfseries\noindent{\abstractname}\par\nobreak}\footnotesize}
  {\bigskip}

\titlespacing{\section}{0pt}{*3}{*1}
\titlespacing{\subsection}{0pt}{*2}{*0.5}
\titlespacing{\subsubsection}{0pt}{*1.5}{0pt}

\usepackage{authblk}

\usepackage{graphicx}
\usepackage[space]{grffile}
\usepackage{latexsym}
\usepackage{textcomp}
\usepackage{longtable}
\usepackage{tabulary}
\usepackage{booktabs,array,multirow}
\usepackage{amsfonts,amsmath,amssymb}
\providecommand\citet{\cite}
\providecommand\citep{\cite}

\newif\iflatexml\latexmlfalse

\AtBeginDocument{\DeclareGraphicsExtensions{.pdf,.PDF,.eps,.EPS,.png,.PNG,.tif,.TIF,.jpg,.JPG,.jpeg,.JPEG}}

\usepackage[utf8]{inputenc}
\usepackage[ngerman,english]{babel}

\usepackage{float}

\usepackage[margin=1.3in]{geometry}

\begin{document}

\title{Dynamical invariants and quantization of the one-dimensional
time-dependent, damped, and driven harmonic oscillator}

\author[1]{M. C. Bertin\footnote{mbertin@ufba.br}}%
\author[1]{J. R. B. Peleteiro}%
\author[2]{B. M. Pimentel}%
\author[1]{J. A. Ramirez}%
\affil[1]{Instituto de Física, Universidade Federal da Bahia}%
\affil[2]{Instituto de Física Teórica, São Paulo State University}%

\vspace{-1em}

  \date{}

\begingroup
\let\center\flushleft
\let\endcenter\endflushleft
\maketitle
\endgroup

\onehalfspacing

\thispagestyle{empty}

\selectlanguage{english}
\begin{abstract}
In this paper, it is proposed a quantization procedure for the
one-dimensional harmonic oscillator with time-dependent frequency,
time-dependent driven force, and time-dependent dissipative term. The
method is based on the construction of dynamical invariants previously
proposed by the authors, in which fundamental importance is given to the
linear invariants of the oscillator.

\end{abstract}%

\sloppy

\begin{quote}
This is a pre-print of an article published in Brazilian Journal of
Physics. The final authenticated version is available online
at:~\href{https://doi.org/\%5Binsert}{https://doi.org/10.1007/s13538-020-00765-8}.
\end{quote}

\section{\texorpdfstring{{Introduction}}{Introduction}}

{\label{404122}}

Dynamical invariants were first used by Ermakov~to show the connection
between solutions of some special differential equations, referred to as
Steen-Ermakov equations~\hyperref[csl:1]{[1]}. These equations were first
studied by Steen~\hyperref[csl:2]{[2]} and then rediscovered by other
authors~\hyperref[csl:3]{[3]}; \hyperref[csl:4]{[4]}. After that, Ray and Reid used the Ermakov
approach to construct invariants for a much broader class of
differential equations~\hyperref[csl:5]{[5]}; \hyperref[csl:6]{[6]}; \hyperref[csl:7]{[7]}. This purely mathematical
interest was the start point of significant developments in classical
and quantum dynamics.

The importance of the dynamical invariants of a system should not be
underrated. In classical mechanics, the dynamical constants of motion
are the variables that allow complete integration of dynamical systems.
In classical field theories, symmetries of lagrangian systems are
related to continuity equations and time-invariants through the Noether
theorem~\hyperref[csl:8]{[8]} In quantum field theory, Casimir invariants
of symmetry groups are essential to the understanding of the fundamental
particle structure of our universe~\hyperref[csl:9]{[9]}.

In quantum mechanics, a complete characterization of a quantum system is
achieved by the knowledge of a complete set of time-invariant
observables, which are also generators of a complete symmetry of the
system. The process of quantization, therefore, is accomplished by
finding an invariant set of stationary eigenvectors which generates,
hopefully, a Hilbert space. Symmetries are linked to invariants, and
invariants are linked to the very existence of quantum states, on a very
fundamental level.

In time-dependent systems, dynamical invariants play a major role, since
the energy is no longer conserved, and sometimes even defined.
Particularly, in quantum mechanics, systems with time-dependent
Hamiltonians do not have well-defined energy spectra. Even in the case
where a complete basis of eigenvectors exists, one cannot be sure that
this condition persists in time. When quantization is allowed,
the~problem of time-dependent hamiltonians can be dealt with by finding
a hermitian quadratic invariant, for which the eigenvalue problem is
well defined~\hyperref[csl:10]{[10]}. Time-dependent systems appear in
several applications in physics such as ion traps~\hyperref[csl:11]{[11]}; \hyperref[csl:12]{[12]}; \hyperref[csl:13]{[13]},
optical cavities~\hyperref[csl:14]{[14]}, and to perform algorithms in
quantum computation~\hyperref[csl:15]{[15]}; \hyperref[csl:16]{[16]}.

There are several methods to calculate dynamical invariants. In the
classical case, we have Lutzky's approach~\hyperref[csl:17]{[17]}; \hyperref[csl:18]{[18]}, which
consists of the application of the Noether theorem. Another method is
the dynamical algebra approach~\hyperref[csl:19]{[19]}; \hyperref[csl:20]{[20]}. Recently, the
authors~developed a new way to calculate dynamical
invariants~\hyperref[csl:21]{[21]}, which consists of combinations of the
equations of motion. These last two methods can be used in both,
classical and quantum cases.

In this work, we show how the definition of first-order invariants
allows us to~approach the quantization of the one-dimension
time-dependent, damped, driven harmonic oscillator (TDDDHO).~In
sec.~{\ref{832125}}, we follow~\hyperref[csl:21]{[21]} and
calculate the linear invariants for the TDDDHO by taking the
combinations of the equations of motion. Next, in
sec.~{\ref{965209}}, we construct the quadratic
invariant and find a Steen-Ermakov-like equation. In
sec.~{\ref{343671}}, we perform the quantization of the
TDDDHO using the algebra of the first-order invariants.
Sec.~{\ref{745680}} presents the coordinate
representation in the form of wave eigenfunctions of the quadratic
invariant, along with a general expression for the uncertainty relations
between the observables~\((q,p)\). In
section~{\ref{943594}}, we address the problem of the
dissipative oscillator with constant parameters and general driven
force. Finally, in sec.~{\ref{549543}}, we present our
main observations.

\section{First-order invariants of the
oscillator}

{\label{832125}}

Let us start with the hamiltonian operator

\begin{equation}
H=\frac{1}{2m}e^{-G\left(t\right)}p^{2}+\frac{1}{2}m\omega^{2}\left(t\right)e^{G\left(t\right)}q^{2}-e^{G\left(t\right)}F\left(t\right)q,\label{eq:01}
\end{equation}

in which the canonical pair~\(\left(q,p\right)\) are Hilbert space
operators with commutation
relations~\(\left[q,p\right]=i\hbar\boldsymbol{1}\),~\(\left[q,q\right]=0\),
and~{\(\left[p,p\right]=0\). The term~\(\omega(t)\)} represents a time
dependent angular frequency,~\(F\left(t\right)\) stands for a time
dependent driven force, and~\(G\left(t\right)\) is another time dependent
function. These functions are supposed to be at least of
class~\(C^{2}\). This operator can be seen as a generalization
of the Bateman-Caldirola-Kanai (BCK) model for the dissipative harmonic
oscillator~\hyperref[csl:22]{[22]}; \hyperref[csl:23]{[23]}; \hyperref[csl:24]{[24]}.

Heisenberg's equations for the hamiltonian
({\ref{eq:01}}) are given by

\begin{subequations}\label{eq:02} 
\begin{align}
 \dot{q}  =&\  e^{-G}p/m,\label{eq:02a}\\
 \dot{p}  =&\  e^{G}F-e^{G}m\omega^{2}q,\label{eq:02b} \\
 \ddot{q} =&\ \frac{1}{m}F-2g\dot{q}-\omega^{2}q,\thinspace\thinspace\thinspace\thinspace\thinspace\thinspace\thinspace g\left(t\right)\equiv\frac{1}{2}\dot{G}\left(t\right).\label{eq:02c}
\end{align}
\end{subequations}

The function~\(g\left(t\right)\) has the interpretation of a dissipative
term.

We proceed by calculating the first-order~dynamical invariants related
to (\ref{eq:02}) with the method proposed
by~\hyperref[csl:21]{[21]}. In this case we define two arbitrary complex
functions~\(\alpha\left(t\right)\) and~\(\beta\left(t\right)\). Multiplying
(\ref{eq:02a}) by~\(\alpha\) and
(\ref{eq:02b}) by~\(\beta\), building the
linear combination, and isolating the total time derivative results in
the expression

\begin{equation}
\frac{d}{dt}\left(\beta p+\alpha mq\right)=\left(\alpha e^{-G}+\dot{\beta}\right)p+m\left(\dot{\alpha}-e^{G}\omega^{2}\beta\right)q+\beta e^{G}F.\label{eq:04}
\end{equation}

Now, we define the function

\begin{equation}
\mathcal{F}\left(\beta,t\right)\equiv\int_{t_{0}}^{t}e^{G\left(\tau\right)}\beta\left(\tau\right)F\left(\tau\right)d\tau,\,\,\,\,\,\,\,\,\,\,\,\,\,\,\,\,\,\,\,\,\beta\left(t_{0}\right)F\left(t_{0}\right)=0,\label{eq:05}
\end{equation}

for which we have the identity

\begin{equation}
\beta e^{G}F=\frac{d\mathcal{F}}{dt}.\label{eq:06}
\end{equation}

With ({\ref{eq:06}}), we may express
({\ref{eq:04}}) in the form

\[
\frac{d}{dt}\left(\beta p+\alpha mq-\mathcal{F}\right)=\left(\alpha e^{-G}+\dot{\beta}\right)p+m\left(\dot{\alpha}-e^{G}\omega^{2}\beta\right)q.\]

If the parameters \(\alpha\) and \(\beta\) satisfy the
ODEs

\begin{align}
 \alpha+e^{G}\frac{d\beta}{dt}&=0,\nonumber\\
 \frac{d\alpha}{dt}-e^{G}\omega^{2}\beta&=0, \nonumber
\end{align}

the polynomial

\begin{equation}
I=\beta p+\alpha mq-\mathcal{F}\left(\beta,t\right)\label{eq:09}
\end{equation}

becomes a first-order invariant of the system
(\ref{eq:02}).

The functions~\(\alpha\) and~\(\beta\) are not
independent solutions, therefore, we may write
({\ref{eq:09}}) depending on~\(\beta\) alone:

\begin{equation}
I=\beta p-me^{G}\dot{\beta}q-\mathcal{F},\label{eq:10}
\end{equation}

where~\(\beta\) is now a solution of the equation

\begin{equation}
\ddot{\beta}+2g\dot{\beta}+\omega^{2}\beta=0.\label{eq:11}
\end{equation}

Now we suppose there is a solution of ({\ref{eq:11}})
with the form

\[\beta \equiv  \rho \left(t\right) e^{i\phi \left(t\right)},\]

with~\(\phi(t)\) and~\(\rho\left(t\right)\) both real functions. Eq.
({\ref{eq:11}}) then becomes

\begin{align}
\ddot{\rho}+2g\dot{\rho}+\left(\omega^{2}-\dot{\phi}^{2}\right)\rho&=0,\nonumber \\
2\dot{\rho}\dot{\phi}+\rho\left(\ddot{\phi}+2g\dot{\phi}\right)&=0, \nonumber
\end{align}

and, in this case, it is straightforward to show that

\[\beta^{*}=\rho\left(t\right)e^{-i\phi\left(t\right)} \]

is also a solution of ({\ref{eq:11}}). Therefore, the
operator

\begin{equation}
I^{\dagger}=\beta^{*}p-me^{G}\dot{\beta}^{*}q-\mathcal{F}^{*}\label{eq:15}
\end{equation}

is also a linearly independent first-order dynamical invariant.

\section{The second-order invariant of the
oscillator}

{\label{965209}}

We may also build quadratic invariants from the equations of motion
(\ref{eq:02}). Without the driving force, it would be
sufficient to build linear combinations of products of these equations.
However, this is not the case when the driving force is in place. Let us
observe the following products between (\ref{eq:02a})
and (\ref{eq:02b}):

\begin{subequations}\label{eq:16}
\begin{align}
\frac{dq^{2}}{dt}&=\frac{e^{-G}}{m}\left\{ q,p\right\} ,\\
\frac{d}{dt}\left\{ q,p\right\} &=2\frac{e^{-G}}{m}p^{2}+2e^{G}Fq-2e^{G}m\omega^{2}q^{2},\\
\frac{dp^{2}}{dt}&=2e^{G}Fp-e^{G}m\omega^{2}\left\{ q,p\right\} ,
\end{align}
\end{subequations}

where~\(\left\{q,p\right\}\equiv qp+pq\) represents the anti-commutator. The r.h.s. of
these equations fail to be purely quadratic forms, because of the
presence of the driving force. This situation is corrected with the use
of the equations of motion~(\ref{eq:02}) themselves.

Now we take a set of time-dependent functions~\(c_i=\left(c_1,c_2,c_3,c_4,c_5\right)\), build
a linear combination of (\ref{eq:02}) and
(\ref{eq:16}), and collect the total time derivative.
The result is given by

\begin{align}
\frac{d}{dt}\left[c_{1}\frac{q^{2}}{2}\right.+ & \left.\frac{1}{2}c_{2}\left\{ q,p\right\} +c_{3}\frac{p^{2}}{2}+c_{4}q+c_{5}p-\mathcal{F}\left(c_{5},t\right)\right]=\nonumber \\
= & \left(\frac{1}{2}\frac{dc_{3}}{dt}+c_{2}\frac{e^{-G}}{m}\right)p^{2}+\left(\frac{1}{2}\frac{dc_{1}}{dt}-c_{2}e^{G}m\omega^{2}\right)q^{2}+\nonumber \\
 & +\frac{1}{2}\left(\frac{dc_{2}}{dt}+c_{1}\frac{e^{-G}}{m}-c_{3}e^{G}m\omega^{2}\right)\left\{ q,p\right\} +\nonumber \\
 & +\left(c_{2}e^{G}F-c_{5}e^{G}m\omega^{2}+\frac{dc_{4}}{dt}\right)q \ +\nonumber \\ &+\left(\frac{dc_{5}}{dt}+c_{4}\frac{e^{-G}}{m}+c_{3}e^{G}F\right)p.\nonumber
\end{align}

Hence, the second-order polynomial

\begin{equation}
I_{Q}=\frac{c_{1}}{2}q^{2}+\frac{c_{2}}{2}\left\{ q,p\right\} +\frac{c_{3}}{2}p^{2}+c_{4}q+c_{5}p-\mathcal{F}\left(c_{5},t\right)\label{eq:18}
\end{equation}

is a dynamical invariant if the equations

\begin{align}
\frac{dc_{3}}{dt}+2\frac{e^{-G}}{m}c_{2}&=0,\nonumber \\
\frac{dc_{1}}{dt}-2c_{2}e^{G}m\omega^{2}&=0,\nonumber \\
\frac{dc_{2}}{dt}+c_{1}\frac{e^{-G}}{m}-c_{3}e^{G}m\omega^{2}&=0, \nonumber \\
\frac{dc_{4}}{dt}+c_{2}e^{G}F-c_{5}e^{G}m\omega^{2}&=0,\nonumber \\
\frac{dc_{5}}{dt}+c_{3}e^{G}F+c_{4}\frac{e^{-G}}{m}&=0, \nonumber
\end{align}

are satisfied.

We notice that (\ref{eq:18}) can be rewritten to depend
only on the functions~\(c_3\) and~\(c_5\). Let us
rename them as~\(\gamma\) and \(\sigma\), respectively.
In this case, the second-order invariant is given by

\begin{align}
I_{Q}  = & \ \frac{1}{2}\left(me^{G}\right)^{2}\left(\frac{1}{2}\frac{d^{2}\gamma}{dt^{2}}+g\frac{d\gamma}{dt}+\omega^{2}\gamma\right)q^{2}-\frac{m}{4}e^{G}\frac{d\gamma}{dt}\left\{ q,p\right\} \nonumber \\
 &   +\frac{\gamma}{2}p^{2}-me^{G}\left(\frac{d\sigma}{dt}+\gamma e^{G}F\right)q+\sigma p-\mathcal{F}\left(\sigma,t\right),\label{eq:20}
\end{align}

and ODEs for~\(\gamma\) and \(\sigma\) follow:

\begin{subequations}\label{eq:21}
\begin{align}
\frac{1}{2}\frac{d^{3}\gamma}{dt^{3}}+3g\frac{d^{2}\gamma}{dt^{2}}+\left(\dot{g}+4g^{2}+2\omega^{2}\right)\frac{d\gamma}{dt}+\left(\frac{d\omega^{2}}{dt}+4\omega^{2}g\right)\gamma=0,\label{eq:21a}\\
\frac{d^{2}\sigma}{dt^{2}}+2g\frac{d\sigma}{dt}+\omega^{2}\sigma=-\frac{3}{2}e^{G}F\frac{d\gamma}{dt}-e^{G}\left(\frac{dF}{dt}+4gF\right)\gamma.\label{eq:21b}
\end{align}
\end{subequations}

Eq. (\ref{eq:21a}) above has a first integral given by

\par\null

\begin{equation}
 \frac{d^2 \gamma}{dt^2} + 2 g \frac{d \gamma}{dt}  + 2 \omega^2 \gamma = \frac{ 1 }{2 \gamma} \left( \frac{d \gamma}{dt} \right)^2 + e^{- 2 G} C, \label{primeiraIntegral}
\end{equation}

which can be turned into a Steen-Ermakov-like equation through the
change of variables~\(\gamma=r^2\):

\par\null

\begin{equation}
\frac{d^2 r}{dt^2} + 2 g r^3 \frac{d r}{dt} + \omega^2 r = \frac{e^{- 2 G} C}{2 r^3}. \label{SteenErmakov}
\end{equation}

The Steen-Ermakov equation itself is obtained when~\(g=0\).
The equation for \(\sigma\) is relevant only if the force term
is present. Otherwise, the above invariants resemble the case of the
oscillator with time-dependent frequency already addressed in the
ref.~\hyperref[csl:21]{[21]}.

\section{Quantization of the
oscillator}

{\label{343671}}

Now we wish to explore the fact that the first-order operators
({\ref{eq:10}}) and ({\ref{eq:15}})
are two dynamical invariants of the oscillator if~\(\beta\)
and~\(\beta^*\) are two L.I. solutions of
({\ref{eq:11}}). The commutation relations are found to
be

\[
\left[I,I^{\dagger}\right] =\Omega\boldsymbol{1}, \ \ \ \Omega\equiv im\hbar e^{G}W,\thinspace\thinspace\textnormal{and}\thinspace\thinspace\thinspace\thinspace\thinspace\thinspace W\equiv\dot{\beta}^{*}\beta-\beta^{*}\dot{\beta}.
\]

The remaining relations are just~\(\left[I,I\right]=\left[I^{\dagger},I^{\dagger}\right]=0\). In fact, using
({\ref{eq:11}}) it is straightforward to see
\(\Omega\) is a constant of motion by itself.

We define the operators

\begin{equation}
a\equiv\frac{I}{\sqrt{\Omega}}, \thinspace\thinspace\thinspace\thinspace\thinspace\thinspace\thinspace\thinspace a^{\dagger}\equiv\frac{I^{\dagger}}{\sqrt{ \Omega}},\label{27}
\end{equation}

which obey the commutation relations

\[
\left[a,a^{\dagger}\right]=\boldsymbol{1},\thinspace\thinspace\thinspace\thinspace\left[a,a\right]=\left[a^{\dagger},a^{\dagger}\right]=0.\]

Since~\(a
\) and~\(a^{\dagger}\) are invariants, any
product between them is also a dynamical invariant. This fact allows the
introduction of the number operator

\begin{equation}
\hat{\boldsymbol{n}} \equiv a^{\dagger}a,\label{29}
\end{equation}

which is a time-conserved self-adjoint quadratic quantity. The
quantization is performed by assuming the existence of a complete set of
eigenstates \(\left|n\right\rangle\), i.e.,

\[
\hat{\boldsymbol{n}}\left|n\right\rangle =n\left|n\right\rangle ,\]

where~\(n\) is a positive real number, because of the
positivity of the inner product.

The complete algebra of the oscillator is shown to be given by

\begin{equation}
\left[a,a^{\dagger}\right]=\boldsymbol{1},\thinspace\thinspace\thinspace\left[\hat{\boldsymbol{n}},a\right]=-a,\thinspace\thinspace\thinspace\left[\hat{\boldsymbol{n}},a^{\dagger}\right]=a^{\dagger},\label{31}
\end{equation}

from where we derive

\begin{equation}
a\left|n\right\rangle =\sqrt{n}\left|n-1\right\rangle ,\thinspace\thinspace\thinspace\thinspace\thinspace\thinspace\thinspace\thinspace\thinspace\thinspace\thinspace a^{\dagger}\left|n\right\rangle =\sqrt{n+1}\left|n+1\right\rangle ,\label{32}
\end{equation}

therefore,~\(a
\) and~\(a^{\dagger}\) are ladder
operators. As usual, we suppose the existence of a fundamental state,
defined by~\(a\left|0\right\rangle =0\), and therefore~\(n\) must be
a natural number. All other eigenstates can be derived from

\begin{equation}
\left|n\right\rangle =\frac{\left(a^{\dagger}\right)^{n}}{\sqrt{n!}}\left|0\right\rangle ,\label{33}
\end{equation}

and the quantization procedure is complete.

We see that the dynamical algebra of the
operators~\(\hat{\boldsymbol{n}}\),~\(a\),
and~\(a^{\dagger}\)is the same as of the simple harmonic oscillator,
so it is the Hilbert space spanned by the~\(\left|n\right\rangle\) states.
What is distinct among the several possible choices of the
parameters~\(\left(g,\omega, \mathcal{F}\right)\) are the behavior of the physical
characteristic functions of the model, as the energy values, expected
values, and others.

\section{Eigenvalue solutions, eigenfunctions, and
uncertainty}

{\label{745680}}

Let us now show the explicit form of the number operator:

\[
\hat{\boldsymbol{n}}=a^{\dagger}a=\frac{1}{\Omega}\left(\frac{1}{2}\left[I^{\dagger},I\right]+\frac{1}{2}\left\{ I^{\dagger},I\right\} \right)=\frac{1}{2}\left(\frac{1}{\Omega}\left\{ I^{\dagger},I\right\}-\boldsymbol{1}\right) .\]

The quantity \(\frac{1}{2}\left\{ I^{\dagger},I\right\}\) is also a quadratic self-adjoint
dynamical invariant, calculated by

\begin{align}
\frac{1}{2}\left\{ I^{\dagger},I\right\}   = & \ \beta^{*}\beta p^{2}-\frac{1}{2}me^{G}\left(\beta^{*}\dot{\beta}+\dot{\beta}^{*}\beta\right)\left\{ q,p\right\} +m^{2}e^{2G}\dot{\beta}^{*}\dot{\beta}q^{2}\nonumber \\
   & -\left(\beta^{*}\mathcal{F}+\mathcal{F}^{*}\beta\right)p+me^{G}\left(\dot{\beta}^{*}\mathcal{F}+\dot{\beta}\mathcal{F}^{*}\right)q+\frac{1}{2}\mathcal{F}^{*}\mathcal{F}.\nonumber
\end{align}

The definition of the real function \(\gamma\equiv2\beta^{*}\beta\) results in the
expression

\begin{align}
\frac{1}{2}\left\{ I^{\dagger},I\right\}   = & \  \frac{1}{2}\left(me^{G}\right)^{2}\left(\frac{1}{2}\frac{d^{2}\gamma}{dt^{2}}+g\frac{d\gamma}{dt}+\omega^{2}\gamma\right)q^{2}-\frac{1}{4}me^{G}\frac{d\gamma}{dt}\left\{ q,p\right\} +\frac{1}{2}\gamma p^{2}\nonumber \\
 &   -\left(\beta^{*}\mathcal{F}+\mathcal{F}^{*}\beta\right)p+me^{G}\left(\dot{\beta}^{*}\mathcal{F}+\mathcal{F}^{*}\dot{\beta}\right)q+\frac{1}{2}\mathcal{F}^{*}\mathcal{F}.\nonumber
\end{align}

Now, we define \(\sigma\equiv-\beta^{*}\mathcal{F}-\mathcal{F}^{*}\beta\), which leads to

\[
\dot{\beta}^{*}\mathcal{F}+\mathcal{F}^{*}\dot{\beta}=-\left(\frac{d\sigma}{dt}+\gamma e^{G}F\right).\]

On the other hand, \(\mathcal{F}^{*}\mathcal{F}=\mathcal{F}\left(\beta^{*},t\right)\mathcal{F}\left(\beta,t\right)=-2\mathcal{F}\left(\sigma,t\right)\). Therefore,

\begin{align}
\frac{1}{2}\left\{ I^{\dagger},I\right\} = & \   \frac{1}{2}\left(me^{G}\right)^{2}\left(\frac{1}{2}\frac{d^{2}\gamma}{dt^{2}}+g\frac{d\gamma}{dt}+\omega^{2}\gamma\right)q^{2}-\frac{1}{4}me^{G}\frac{d\gamma}{dt}\left\{ q,p\right\} +\frac{1}{2}\gamma p^{2}\nonumber \\
 &  \  -me^{G}\left(\frac{d\sigma}{dt}+\gamma e^{G}F\right)q+\sigma p-\mathcal{F}\left(\sigma,t\right),\nonumber
\end{align}

which is precisely the second-order invariant
({\ref{eq:20}}). The above result implies

\begin{equation}
I_{Q}=\Omega\left(\hat{\boldsymbol{n}}+\frac{1}{2}\right),\label{39}
\end{equation}

so~\(I_{Q}\) has the same eigenstates of~\(\hat{\boldsymbol{n}}\).

Moreover, considering~\(\left\langle q\left|a\right|0\right\rangle =0\), and the eigenvalue problem
\(q\left|q'\right>=q'\left|q'\right>\), the eigenfunction of the fundamental state obeys the
equation

\[
\left(\mathcal{F}+me^{G}\dot{\beta}q+i\hbar\beta\frac{d}{dq}\right)\psi_{0}\left(q\right)=0,\]

which has the solution

\begin{equation}\psi_{0}=A\exp\left[-\frac{1}{2}\frac{1}{i\hbar\beta}\left(me^{G}\dot{\beta}q^{2}+2\mathcal{F}q\right)\right], \label{26a} \end{equation}

with the normalization constant

\[A=\left(\frac{1}{2\pi\hbar^{2}}\frac{\Omega}{\beta^{*}\beta}\right)^{1/4}\exp\left[-\left(\frac{1}{\beta^{*}\beta}\right)^{2}\frac{1}{\Omega}\left(\mathrm{Im}\left(\beta^{*}\mathcal{F}\right)\right)^{2}\right].\]

The complete set of normalized eigenfunctions are found to be

\begin{equation}\label{n-functions}\psi_{n}=\frac{1}{\sqrt{2^{n}\cdot n!}}\left(i\sqrt{\frac{\beta^{*}}{\beta}}\right)^{n}\psi_{0}H_{n}\left[\sqrt{\frac{\Omega}{2\beta^{*}\beta}}\left(\frac{q}{\hbar}+\frac{2}{\Omega}\mathrm{Im}\left(\beta^{*}\mathcal{F}\right)\right)\right],\end{equation}

where~\(H_n(x)\) are the Hermite polynomials. Here, we stress
the fact that ({\ref{n-functions}}) are eigenfunctions
of the operator~\(I_Q\), but they are also solutions of the
Schr\selectlanguage{ngerman}ödinger equation~\(\left(i\hbar \partial_t -H\right)\psi=0\). These states are the same found
in~\hyperref[csl:25]{[25]}, where coherent states of the general
one-dimensional oscillator are discussed.

Writing the canonical variables in the form

\begin{align}q&=\frac{i\hbar}{\sqrt{\Omega}}\left(\beta^{\ast}a-\beta a^{\dagger}\right)-2\frac{\hbar}{\Omega}\mathrm{Im}\left(\beta^{\ast}\mathcal{F}\right), \nonumber \\p&=\frac{im\hbar e^{G}}{\sqrt{\Omega}}\left(\dot{\beta}^{\ast}a-\dot{\beta}a^{\dagger}\right)-\frac{2m\hbar e^{G}}{\Omega}\mathrm{Im}\left(\dot{\beta}^{\ast}\mathcal{F}\right) \nonumber,\end{align}

allows us to calculate the uncertainty relations

\[\left(\Delta q\right)^{2}_n\left(\Delta p\right)^{2}_n=\frac{2m^{2}\hbar^{4}e^{2G}}{\Omega^{2}}\gamma\dot{\beta}^{*}\dot{\beta}\left(n+\frac{1}{2}\right)^{2}.\]

\section{The underdamping oscillator}

{\label{943594}}

Let us analyze the case~\(g^2 \le \omega^2\) with both~\(\omega\)
and~\(g\) constant parameters, and~\(F=F(t)\) still
arbitrary. In this case, the function~\(G\) should be
linear in~\(t\). Let us suppose it to have the form
of~\(G=2gt\). We also have the solution

\begin{equation}\beta = \exp(-gt)\exp(i \bar{\omega}t), \ \ \ \ \ \bar{\omega}^2\equiv\omega^2-g^2,\label{49}\end{equation}

while~\(\beta^*\) is just the complex conjugate. With
({\ref{49}}), the linear dynamical invariants of the
system become

\[I=   e^{i\bar{\omega}t}\left[e^{-gt}p+m\left(g-i\bar{\omega}\right)e^{gt}q-e^{-i\bar{\omega}t}\mathcal{F}\right],\]

together with the adjoint operator~\(I^{\dagger}\). We also have the
function{{~}}\(\Omega=2m\bar{\omega}\hbar\){, }which gives the ladder operators

\[a=\frac{e^{i\bar{\omega}t}}{\sqrt{2m\bar{\omega}\hbar}}\left[e^{-gt}p+m\left(g-i\bar{\omega}\right)e^{gt}q-e^{-i\bar{\omega}t}\mathcal{F}\right],\]

and the adjoint \(a^{\dagger}\).

In this case, the quadratic invariant can be found from
({\ref{eq:20}}):

\par\null

\begin{align}
I_{Q}=  & \ e^{-2gt}p^{2}+m^{2}\omega^{2}e^{2gt}q^{2}+mg\left\{ q,p\right\}\nonumber \\& \ -me^{G}\left(\frac{d\sigma}{dt}+\gamma e^{G}F\right)q+\sigma p-\mathcal{F}\left(\sigma,t\right).\label{32b}
\end{align}

Note that~\(I_Q\) is an invariant observable, so the
invariant eigenvalues

\begin{equation}
I_n = 2 m \overline{\omega}\hbar\left(n + \frac{1}{2}\right)\label{eq:53}
\end{equation}

represent invariant characteristic values of the oscillator.

It is possible to calculate the fundamental eigenfunction with the use
of ({\ref{26a}}), resulting in the normalized function

\begin{align}\psi_{0}=& \ e^{gt/2}\left(\frac{m\bar{\omega}}{\pi\hbar}\right)^{1/4}\exp\left[-\frac{1}{2m\hbar\bar{\omega}}e^{2gt}\left(\mathrm{Im}\left(e^{-i\bar{\omega}t}\mathcal{F}\right)\right)^{2}\right]\times \nonumber \\ &
\ \times \exp\left(-\frac{i}{2}\frac{m\bar{g}}{\hbar}e^{2gt}q^{2}+\frac{i}{\hbar}e^{-i\bar{\omega}t}\mathcal{F}e^{gt}q\right) \label{33l} \end{align}

where \(\overline{g}\equiv g-i\overline{\omega}\). A straightforward calculation shows that

\begin{equation}\psi_{n}=\frac{\left(i\right)^{n}}{\sqrt{2^{n}\cdot n!}}e^{-in\bar{\omega}t}\psi_{0}H_{n}\left(x\right)\label{34l}\end{equation}

are the normalized eigenfunctions, where

\[x=\sqrt{\frac{m\bar{\omega}}{\hbar}}e^{gt}q-\sqrt{\frac{1}{m\hbar\bar{\omega}}}\mathrm{Im}\left(e^{-i\bar{\omega}t}\mathcal{F}\right).\]

Moreover, we have the uncertainty relations

\[\left(\Delta q\right)_{n}^{2}\left(\Delta p\right)_{n}^{2}=\hbar^{2}\frac{\bar{\omega}^{2}-g^{2}}{\bar{\omega}^{2}}\left(n+\frac{1}{2}\right)^{2},\]

which are~time-independent.

Let us display some results for the case~\(F=F_0\sin\left(\alpha t\right)\). Since the
force term does not change the differential equations
for~\(\beta\) and\(\ \beta^{\ast}\), their solutions are the
same as the ones proposed in this section. Now that we have a force term
we need to calculate~\(\mathcal{F}\) given by
({\ref{eq:05}}),~resulting

\[
\mathcal{F}=F_{0}e^{\bar{g}^{*}t}\frac{\overline{g}^{*}\sin\left(\alpha t\right)-\alpha\cos\left(\alpha t\right)+\alpha e^{-\bar{g}^{*}t}}{\left(\overline{g}^{*}\right)^{2}+\alpha^{2}}. \]

The parameter~\(\sigma\) is given by

\begin{align}
\sigma=& \ \frac{-2F_{0}}{\left[\left(g^{2}-\overline{\omega}^{2}+\alpha^{2}\right)^{2}+4\overline{\omega}^{2}g^{2}\right]} \left\{ 2\overline{\omega}g\left[\overline{\omega}\sin(\alpha t)-\alpha e^{-gt}\sin(\overline{\omega}t)\right] \right. \nonumber \\ & \left. + \left(g^{2}-\overline{\omega}^{2}+\alpha^{2}\right)\left[g\sin(\alpha t)-\alpha\cos(\alpha t)+\alpha e^{-gt}\cos(\overline{\omega}t)\right] \right\} \nonumber,
\end{align}

and~\(\mathcal{F} (\sigma,t)\)~becomes

\begin{align}
\mathcal{F}\left(\sigma,t\right)=& \  \frac{-2F_{0}^{2}e^{2gt}}{\left[\left(g^{2}-\overline{\omega}^{2}+\alpha^{2}\right)^{2}+4\overline{\omega}^{2}g^{2}\right]} \left\{ \left[\overline{\omega}\sin(\alpha t) -\alpha e^{-gt}\sin(\overline{\omega}t)\right]^{2} \right. \nonumber \\ & \left. +\left[g\sin(\alpha t) -\alpha \cos(\alpha t) +\alpha e^{-gt}\cos(\overline{\omega}t)\right]^{2} \right\} \nonumber.
\end{align}

The simple harmonic oscillator is trivially recovered in the
case~\(G=0\),~\(\overline{\omega}=\omega\), and~\(\mathcal{F}=0\),
which also gives the condition~\(\sigma=0\).

\section{\texorpdfstring{{Further
observations}}{Further observations}}

{\label{549543}}

In this work, we showed a procedure for the quantization of the harmonic
oscillator with time-dependent frequency, time-dependent driven force,
and time-dependent dissipative term. The procedure is based~on the
construction of the linear invariants of the BCK Hamiltonian
({\ref{eq:01}}), which turns out to be ladder
operators. We also construct the Hilbert space of the system and
calculate the wave eigenfunctions.

This approach shows that the fundamental quantities turn out to be the
linear invariants. Other attempts of analyzing the quantum oscillator
from the dynamical invariant point of view can be found in the
literature, most of them are based on the second-order invariant
({\ref{27}}) as the proper Hamiltonian operator, as the
case of~\hyperref[csl:26]{[26]}. However, the fundamental role of the linear
invariants for the quantization of the oscillators can be found
in~\hyperref[csl:27]{[27]}; \hyperref[csl:25]{[25]}. In fact, the procedure of the
ref.~\hyperref[csl:25]{[25]}~is very close to the one employed here. In the
case of the underdamped oscillator, we also report to the
refs.~\hyperref[csl:28]{[28]}, where the authors propose a quantization
procedure based on the construction of first-order actions, and also to
the ref.~\hyperref[csl:29]{[29]}.

We found that the abstract Hilbert space of the general quadratic
oscillator is the same as the simple harmonic oscillator. However, it is
not a surprise that the same is not observed with the solutions of the
Schr\selectlanguage{ngerman}ödinger equation, which are also eigenfunctions of the quadratic
invariant~\(I_Q\). The wave functions~\(\psi_n\) are
time-dependent and lead, in the general case, to time-dependent
expectation values and uncertainty relations for the canonical
operators. In the special case of constant parameters, however,
the~uncertainty relations between~\(q\)
and~\(p\) are time-independent.

We note that the procedure in~\hyperref[csl:21]{[21]} does not need a
Hamiltonian function, but can be implemented from the equation of motion
(\ref{eq:02c}). However, some caution would be advised.
First, the first-order equations,

\[\dot{x}=y,\ \ \ \ \ \ \dot{y}=-\dot{G}y-\omega^{2}x+F/m,\]

do not constitute a canonical system, since it is not compatible with
the condition~\(\left[x,y\right]=i\hbar\). A direct calculation shows that

\[\frac{d}{dt}\left[x,y\right]=-\dot{G}\left[x,y\right],\ \ \ \ \implies\ \ \ \ \left[x,y\right]=i\hbar\exp\left(-\dot{G}t\right).\]

This result alone would make us believe that the system is indeed
dissipative since it is clear that the allowed classical states would
collapse to the zero volume in time. However, if there would be a local
transformation to a set of canonical variables, a volume preserved
phase-space would emerge. This phase-space would obey the Darboux and
the Liouville theorems. The condition for the existence of such
transformation is given by~\(\left\{ x,y\right\} =e^{-\dot{G}t}\), where~\(\left\{\bullet,\bullet\right\}\)
are the Poisson brackets with respect to the
variables~\(\left(q,p\right)\). This condition is indeed quite general.
However, the only allowed transformation that leads to the two first
equations of ({\ref{eq:02a}}) is given
by~\(x=q\), and~\(y=e^{-\dot{G}t/2}p\)
provided~\(G\) is homogeneous of degree zero. Both sets of
first-order equations are not generally compatible.

\subsection*{Acknowledgments}

This study was financed in part by the Coordenação de Aperfeiçoamento de
Pessoal de Nível Superior -- Brasil (CAPES) -- Finance Code 001. B. M.
Pimentel thanks the Conselho Nacional de Desenvolvimento Científico e
Tecnológico (CNPq) for partial support.

\selectlanguage{english}
\FloatBarrier
\section*{References}\sloppy
\phantomsection
\label{csl:1}1. P. Ermakov, Applicable Analysis and Discrete Mathematics \textbf{2}, 123 (2008).

\phantomsection
\label{csl:2}2. R. Redheffer, Aequationes Mathematicae \textbf{61}, 131 (2001).

\phantomsection
\label{csl:3}3. W. E. Milne, Physical Review \textbf{35}, 863 (1930).

\phantomsection
\label{csl:4}4. E. Pinney, Proceedings of the American Mathematical Society \textbf{1}, 681 (1950).

\phantomsection
\label{csl:5}5. J. R. Ray and J. L. Reid, Physics Letters A \textbf{71}, 317 (1979).

\phantomsection
\label{csl:6}6. J. R. Ray and J. L. Reid, Physics Letters A \textbf{74}, 23 (1979).

\phantomsection
\label{csl:7}7. J. R. Ray, Physics Letters A \textbf{78}, 4 (1980).

\phantomsection
\label{csl:8}8. D. E. Neuenschwander, \textit{{Emmy Noether's Wonderful Theorem}} (Johns Hopkins University Press, 2011).

\phantomsection
\label{csl:9}9. E. Wigner, The Annals of Mathematics \textbf{40}, 149 (1939).

\phantomsection
\label{csl:10}10. H. R. Lewis and W. B. Riesenfeld, Journal of Mathematical Physics \textbf{10}, 1458 (1969).

\phantomsection
\label{csl:11}11. W. Paul, Reviews of Modern Physics \textbf{62}, 531 (1990).

\phantomsection
\label{csl:12}12. D. Leibfried, R. Blatt, C. Monroe, and D. Wineland, Reviews of Modern Physics \textbf{75}, 281 (2003).

\phantomsection
\label{csl:13}13. E. Torrontegui, S. Ib{\'{a}}{\~{n}}ez, X. Chen, A. Ruschhaupt, D. Gu{\'{e}}ry-Odelin, and J. G. Muga, Physical Review A \textbf{83}, (2011).

\phantomsection
\label{csl:14}14. H. Johnston and S. Sarkar, Journal of Physics A: Mathematical and General \textbf{29}, 1741 (1996).

\phantomsection
\label{csl:15}15. M. S. Sarandy, E. I. Duzzioni, and R. M. Serra, Physics Letters A \textbf{375}, 3343 (2011).

\phantomsection
\label{csl:16}16. U. Güngördü, Y. Wan, M. A. Fasihi, and M. Nakahara, Physical Review A \textbf{86}, (2012).

\phantomsection
\label{csl:17}17. M. Lutzky, Journal of Physics A: Mathematical and General \textbf{11}, 249 (1978).

\phantomsection
\label{csl:18}18. M. Lutzky, Physics Letters A \textbf{68}, 3 (1978).

\phantomsection
\label{csl:19}19. H. J. Korsch, Physics Letters A \textbf{74}, 294 (1979).

\phantomsection
\label{csl:20}20. R. S. Kaushal and H. J. Korsch, Journal of Mathematical Physics \textbf{22}, 1904 (1981).

\phantomsection
\label{csl:21}21. M. C. Bertin, B. M. Pimentel, and J. A. Ramirez, Journal of Mathematical Physics \textbf{53}, 042104 (2012).

\phantomsection
\label{csl:22}22. H. Bateman, Physical Review \textbf{38}, 815 (1931).

\phantomsection
\label{csl:23}23. P. Caldirola, Il Nuovo Cimento \textbf{18}, 393 (1941).

\phantomsection
\label{csl:24}24. E. Kanai, Progress of Theoretical Physics \textbf{3}, 440 (1948).

\phantomsection
\label{csl:25}25. V. V. Dodonov and V. I. Man{\textquotesingle}ko, Physical Review A \textbf{20}, 550 (1979).

\phantomsection
\label{csl:26}26. H. R. Lewis, Physical Review Letters \textbf{18}, 636 (1967).

\phantomsection
\label{csl:27}27. I. A. Malkin and V. I. Man{\textquotesingle}ko, Physics Letters A \textbf{32}, 243 (1970).

\phantomsection
\label{csl:28}28. D. M. Gitman and V. G. Kupriyanov, The European Physical Journal C \textbf{50}, 691 (2007).

\phantomsection
\label{csl:29}29. M. C. Baldiotti, R. Fresneda, and D. M. Gitman, Physics Letters A \textbf{375}, 1630 (2011).

\end{document}